\documentclass[journal]{IEEEtran}
\IEEEoverridecommandlockouts
\usepackage{cite}
\usepackage{amsmath,amssymb,amsfonts}

\usepackage{textcomp}
\usepackage{xcolor}
\usepackage{multirow}
\usepackage{makecell}
\usepackage{algorithm,algcompatible}
\usepackage{subfigure}
\usepackage{graphicx}
\usepackage{array}
\usepackage{algpseudocode}
\newcolumntype{P}[1]{>{\centering\arraybackslash}p{#1}}
\usepackage{calc}
\usepackage{array}
\usepackage{optidef}
\usepackage{epstopdf}
\usepackage{mathtools}
\usepackage{mathrsfs}
\usepackage{bm}
\usepackage{cite}
\usepackage{boldline}
\usepackage{algorithm}
\usepackage{algpseudocode}

\usepackage{booktabs}

\newtheorem{definition}{Definition}

\usepackage{siunitx}
\def\BibTeX{{\rm B\kern-.05em{\sc i\kern-.025em b}\kern-.08em
    T\kern-.1667em\lower.7ex\hbox{E}\kern-.125emX}}

\long\def\comment#1{}

\newfont{\bbb}{msbm10 scaled 700}

\newfont{\bb}{msbm10 scaled 1100}
\newcommand{\CC}{\mbox{\bb C}}

\newcommand{\EE}{\mbox{\bb E}}

\newcommand{\av}{{\bf a}}

\newcommand{\hv}{{\bf h}}

\newcommand{\pv}{{\bf p}}

\newcommand{\rv}{{\bf r}}
\newcommand{\sv}{{\bf s}}

\newcommand{\wv}{{\bf w}}

\newcommand{\xv}{{\bf x}}
\newcommand{\yv}{{\bf y}}
\newcommand{\zv}{{\bf z}}

\newcommand{\Dm}{{\bf D}}

\newcommand{\Fm}{{\bf F}}

\newcommand{\Hm}{{\bf H}}
\newcommand{\Id}{{\bf I}}

\newcommand{\Um}{{\bf U}}
\newcommand{\Wm}{{\bf W}}

\newcommand{\Xm}{{\bf X}}

\newcommand{\Cc}{{\cal C}}

\newcommand{\Nc}{{\cal N}}

\newcommand{\SNR}{{\sf SNR}}

\newcommand{\eqdef}{\stackrel{\Delta}{=}}

\newcommand{\herm}{{\sf H}}

\newtheorem{remark}{Remark}

\begin{document}

\title{CSIT-Free Beamforming for Multi-Group Multicast in Overloaded mmWave Systems}
\author{Wonseok Choi,~\IEEEmembership{Student Member,~IEEE}, Jeongjae~Lee,~\IEEEmembership{Student Member,~IEEE},        and~Songnam~Hong,~\IEEEmembership{Senior Member,~IEEE}
\thanks{W. Choi, J. Lee, and S. Hong are with the Department of Electronic Engineering, Hanyang University, Seoul, Korea (e-mail: \{ryan4975, lyjcje7466, snhong\}@hanyang.ac.kr).}

\thanks{This work was supported in part by the Institute of Information \& communications Technology Planning \& Evaluation (IITP) under the artificial intelligence semiconductor support program to nurture the best talents (IITP-2025-RS-2023-00253914) grant funded by the Korea government(MSIT) and in part by the National Research Foundation of Korea(NRF) grant funded by the Korea government(MSIT)(No. RS-2024-00409492).}
}

\maketitle

\begin{abstract}
We study downlink multi-group multicast (MGM) transmission in {\em overloaded} millimeter-wave (mmWave) systems, where the number of users exceeds the number of transmit antennas. We first show that, under realistic line-of-sight (LoS)-dominant user geometries, the conventional single-slot MGM scheme suffers from a fundamental collapse of the max-min fairness degrees of freedom (MMF-DoF), regardless of beamforming optimization. Although this collapse can in principle be avoided via aggressive time-division scheduling, it requires excessive time sharing and results in severe throughput loss in overloaded regimes. To address this limitation, we propose a CSIT-free multi-group multicast framework (CF–MGM) that does not rely on instantaneous channel state information at the transmitter (CSIT) and is based on a deterministic multi-slot transmission structure. By exploiting structured precoding and receiver-side combining across multiple slots, the proposed framework eliminates inter-group interference by construction. We show that CF–MGM guarantees a strictly positive MMF–DoF in overloaded LoS mmWave systems, in sharp contrast to the DoF collapse of conventional single-slot MGM. Simulation results demonstrate that CF–MGM significantly outperforms state-of-the-art CSIT-based MGM schemes while substantially reducing signaling overhead.
\end{abstract}


\section{Introduction}\label{sec:1}
Wireless communication systems continue to evolve to meet the ever-increasing demand for higher data rates, improved energy efficiency, and lower latency. Among emerging communication paradigms, {\em wireless multicasting} has attracted considerable attention as an efficient solution for large-scale content distribution and is widely recognized as a key enabler of future wireless traffic patterns. By allowing the simultaneous delivery of common data to multiple users, multicast transmission can achieve substantial gains in spectral efficiency compared to conventional unicast-based approaches. The importance of multicast communication is further amplified by the rapid proliferation of dense Internet of Things (IoT) deployments and data-centric wireless services~\cite{10263616,7853753,9037357}. In such scenarios, a large number of devices often request identical or highly correlated content, leading to severely overloaded network conditions in which the number of active users far exceeds the available radio resources. These characteristics pose fundamental challenges to traditional downlink transmission strategies that are primarily designed to serve users individually.

In this context, {\em multi-group multicast} (MGM) has emerged as a key enabling technology for scalable wireless networks \cite{Zhu2018,Dong2020}. In MGM, users are partitioned into multiple groups according to their content demands, and each group is served by a dedicated multicast stream. By exploiting shared data interests among users, MGM significantly improves spectral efficiency while reducing signaling overhead, making it particularly attractive for large-scale IoT networks and content-centric services that require massive connectivity and efficient resource utilization.

At the same time, to access substantially larger spectrum resources, operation in high-frequency bands—especially the millimeter-wave (mmWave) spectrum—has become increasingly important for meeting future throughput requirements \cite{Wang2018}. Consequently, supporting multicasting in mmWave systems is of significant practical interest, particularly for content-centric and group-oriented services. However, communication at mmWave frequencies faces several inherent challenges. Severe path loss and high sensitivity to blockage fundamentally limit the link budget, necessitating the use of large antenna arrays and highly directional beamforming to achieve sufficient array gain~\cite{giordani2018tutorial,gapeyenko2017temporal}. Moreover, the deployment of fully digital large-scale antenna arrays is often impractical due to prohibitive hardware cost and power consumption, as each antenna element typically requires a dedicated radio-frequency (RF) chain. Consequently, practical mmWave systems commonly adopt architectures with a limited number of RF chains, such as hybrid analog-digital beamforming. This hardware constraint effectively restricts the available spatial degrees of freedom (DoF) for beamforming and interference suppression, which becomes particularly problematic in dense and overloaded network scenarios. 

\subsection{Challenges}

In MGM, achieving directional beamforming gain is inherently more challenging than in unicast transmission. Unlike the unicast case, a multicast beam must simultaneously serve multiple users whose angles of departure (AoDs) and path gains generally differ. To ensure intra-group fairness, the beamformer cannot simply concentrate power along a single dominant direction; instead, it must balance coverage of the group's angular spread against the suppression of signal leakage toward other groups. This tradeoff becomes particularly pronounced in overloaded, line-of-sight (LoS)-dominant regimes, which are characteristic of mmWave systems. In such scenarios, users within each multicast group span a multi-dimensional angular subspace, rendering system performance highly sensitive to user geometry and imposing fundamental limitations on conventional beamforming strategies.

The performance of MGM is fundamentally governed by the geometry of the underlying channel subspaces, which is characterized by (i) the intra-group channel alignment and (ii) the level of inter-group subspace separation. Under independent and identically distributed (i.i.d.) Rayleigh fading, user channels within each group are typically weakly correlated, making it difficult for a single multicast beam to equalize user rates. Simultaneously, the inherent channel randomness often induces near-orthogonality across different groups, which facilitates inter-group interference suppression. In contrast, under {\em LoS-dominant} conditions, users with similar AoD tend to concentrate their signal energy along the same steering direction, causing each group's channel subspace to effectively collapse into a low-dimensional LoS subspace. While such strong intra-group angular alignment can be beneficial—allowing a single beam to efficiently serve all users within a group—it simultaneously places stringent requirements on inter-group angular separation. When the LoS directions of different groups are insufficiently separated, their channel subspaces become highly correlated, resulting in severe inter-group interference and a pronounced degradation in MGM performance.

In practical deployments, user locations are largely random, making it difficult to guarantee sufficient inter-group angular separation, particularly in overloaded systems~\cite{10263616,lyu2024rate}. Consequently, max-min fairness (MMF) guarantees become fragile. In some channel realizations, multicast groups are well separated in angle, yielding high multicast efficiency; in others, users from multiple groups cluster within similar angular sectors, leading to strong inter-group interference and severe performance degradation—even at high signal-to-noise ratio (SNR) levels. This fragility is further exacerbated under strong LoS conditions, as LoS-dominant propagation shrinks the effective channel subspaces and amplifies cross-group correlation whenever angular separation is limited. In millimeter-wave (mmWave) communications, where propagation is typically sparse and LoS-dominant, narrow beams are effective only when user groups occupy sufficiently distinct angular sectors. However, random user placement often leads to inter-sector overlap, undermining beam orthogonality. In such cases, conventional suboptimal algorithms~\cite{chang2008approximation,tran2013conic,fang2025optimal}—such as semidefinite relaxation (SDR) and successive convex approximation (SCA)—struggle to satisfy MMF constraints without increasing excessive power consumption or complex scheduling overhead. These observations highlight a key insight: mmWave MGM is inherently {\em geometry-limited}, in the sense that its performance depends primarily on effectively exploiting inter-group angular separability, rather than merely increasing transmit power or the number of antennas.


Recent studies have investigated rate-splitting multiple access (RSMA) as a promising approach to mitigate the interference-limited performance of overloaded MGM systems~\cite{joudeh2017rate,yin2020rate,dizdar2023rsma}. By introducing a common (or super-common) stream that can be decoded by multiple multicast groups, RSMA allows part of the inter-group interference to be transformed into a decodable signal component. After successive interference cancellation (SIC), the residual interference experienced by each group stream can be reduced. Consequently, RSMA can improve the worst-case user SINR and fairness under a geometry-limited condition where angular overlap makes interference suppression by conventional single-layer MGM precoding ineffective.


Despite these benefits, RSMA does not eliminate the fundamental geometry-limited nature of mmWave MGM, especially in LoS-dominant dense deployments. Since the common stream must be decoded by all users in its decoding set, its achievable rate is inherently dictated by the weakest effective user across the decoding set; when the AoDs of different multicast groups are not sufficiently separated, worst-user constraint can become tight and restrict how much interference can be efficiently shifted into the common layer. Consequently, even after SIC, significant residual inter-group interference may persist in overloaded and LoS-dominant scenarios. This reveals an important limitation: mmWave MGM is inherently {\em geometry-limited}, meaning that its performance is primarily governed by inter-group angular separability rather than by transmit power or antenna array size alone. As a result, CSIT-based interference management techniques—including RSMA—cannot fully eliminate this bottleneck in dense mmWave deployments.


Beyond this geometry-induced limitation, the heavy reliance on accurate and timely channel state information at the transmitter (CSIT) introduces an additional and often dominant practical challenge. In mmWave systems, acquiring CSIT typically requires substantial beam management and beam training overhead for initial access, beam alignment, and tracking, as well as channel estimation and feedback procedures whose cost increases with both the antenna-array size and the number of users~\cite{hassan2020channel,alkhateeb2014channel,giordani2018tutorial}. These overheads become particularly severe in overloaded scenarios, where many users must be trained and fed back using high-dimensional codebooks. Moreover, user mobility and blockage shorten the effective channel coherence time, forcing more frequent re-training and further reducing the resources available for data transmission.

Taken together, these observations highlight the need for new MGM transmission paradigms that can robustly cope with unfavorable user geometry while significantly reducing the reliance on instantaneous CSIT. Such designs are essential for achieving scalable fairness and reliable performance in overloaded mmWave systems.

\subsection{Contributions}

In this paper, we identify fundamental limitations of existing MGM schemes in overloaded mmWave systems and introduce a CSIT-free MGM framework that effectively overcomes these limitations. The main contributions of this paper are summarized as follows.

\begin{itemize}

\item \textbf{Geometry-limited nature and MMF--DoF collapse of conventional MGM.} We revisit overloaded mmWave MGM from a geometric perspective and show that the feasibility and robustness of MMF are critically determined by the relative AoDs among user groups. Under LoS-dominant propagation conditions typical of mmWave systems, conventional CSIT-based MMF beamforming becomes inherently interference-limited. As a result, the worst-case user rate saturates in the high SNR regime, leading to a fundamental collapse of the MMF--DoF. 

\item \textbf{CSIT-free CF--MGM transmission architecture.} We propose a CSIT-free multi-group multicast (CF--MGM) framework based on a structured multi-slot deterministic transmission using circulant-permutation discrete Fourier transform (CP-DFT) precoding and simple receiver-side linear combining, proposed in \cite{lee2025blind}. Unlike conventional directional beamforming schemes that rely on instantaneous CSIT and favorable inter-group angular separation, the proposed framework eliminates inter-group interference {\em by construction} across slots, while requiring only lightweight channel state information at the receiver (CSIR).

\item \textbf{Low-dimensional formulation and closed-form optimal power allocation.} By exploiting the CF--MGM structure, we show that the original MMF optimization problem can be reduced to a low-dimensional power-allocation problem with a convex feasible set. A closed-form optimal power allocation strategy is derived under the MMF criterion, achieving linear computational complexity and completely avoiding the use of generic convex solvers such as CVX.

\item \textbf{Performance validation under realistic mmWave channels.} Extensive numerical simulations under practical LoS-dominant mmWave channels demonstrate that the proposed CF--MGM framework significantly outperforms optimization-based and CSIT-dependent benchmark schemes, including SDR-, SCA-, HFPI-, PSA-, and MRT-based methods. Notably, CF--MGM achieves substantially higher MMF rates without relying on instantaneous CSIT, while also offering markedly improved computational efficiency. Moreover, unlike conventional CSIT-based MGM approaches, the proposed CSIT-free framework effectively avoids the high-SNR performance saturation (i.e., MMF--DoF collapse) caused by geometry-induced inter-group interference.

\end{itemize}

\subsection{Organization}

The remainder of this paper is organized as follows. Section II describes the system model and introduces the considered multi-group multicast framework. Section III analyzes the fundamental limitations of conventional single-slot MGM and formulates the time-division MGM problem. Section IV presents the proposed CSIT-free CF–MGM framework, including its transmission structure, power allocation, and DoF analysis.  Section VI provides numerical simulation results to validate the performance of CF–MGM. Finally, Section VII concludes the paper.

{\em Notations.} For any positive integer $N$, let $[N]\eqdef \{1,2,...,N\}$. Given an $M\times N$ matrix $\Xm$, $\Xm(i,:)$ and $\Xm(:,j)$ denote its $i$-th row and $j$-th column of $\Xm$, respectively. The operators $(\cdot)^{*}$, $(\cdot)^{\mathsf T}$ and $(\cdot)^{\herm}$ denote element-wise complex conjugate, transpose and Hermitian transpose, respectively. The set $\mathbb{R}_+$ denotes the nonnegative real numbers, i.e., $\mathbb{R}_+ \triangleq \{x \in \mathbb{R} \mid x \ge 0\}$. The notation $\mbox{diag}(\xv)$ represents an $M\times M$ diagonal matrix whose $m$-th diagonal entry equals the $m$-th element of $\xv$ for $m\in[M]$. Finally, $\Id_M$ denotes the $M\times M$ identity matrix.

\section{Preliminaries}\label{sec:2}%

We present the system model and describe the multi-slot transmission frame structure adopted throughout the paper.

\subsection{System Model}\label{subsec:system}
We consider a downlink multi-group multicast (MGM) system operating in a millimeter-wave (mmWave) single-cell network. A base station (BS) equipped with $N$ antennas simultaneously serves $G$ non-overlapping multicast groups. User grouping is assumed to be predetermined by the multicast service, such that users requesting the same content are assigned to the same group. In each group $g\in[G]$, there are $K_g$ single-antenna users randomly distributed within the cell. The total number of users in the system is $K=\sum_{g=1}^{G}K_g$. All users in group $g$ request a common multicast data stream $s_g\in\mathbb C$. We focus on an {\em overloaded} MGM regime in which the total number of users is much larger than the number of antennas at the BS, i.e., $K\gg N$. Nevertheless, the BS is assumed to be equipped with a sufficient number of antennas to support the simultaneous transmission of $G$ independent multicast streams, i.e., $G\le N$.

{
To characterize the mmWave propagation environment, we adopt the widely used Rician fading channel model. The channel vector between the BS and user $k\in[K_g]$ in group $g$ is given by
\begin{equation}\label{eq:channelmodel}
    \hv_{g,k} = \sqrt{\frac{\kappa}{\kappa+1}}\hv_{g,k}^{\rm LoS} + \sqrt{\frac{1}{\kappa+1}}\hv_{g,k}^{\rm NLoS} \in \CC^{N}.
\end{equation} where $\kappa>0$ denotes the Rician $\kappa$-factor. Following standard mmWave large-scale fading modeling, the average attenuation is captured by a log-distance path-loss model characterized by the free-space path loss (FSPL) at a reference distance $d_0$, with an additional log-normal shadowing term~\cite{hemadeh2017millimeter}. In particular, 
\begin{equation}\label{eq:PL_ref}
    PL(d)\,[{\rm dB}] = \Big( PL_0 + 10 n_p \log_{10}\!\frac{d}{d_0} \Big) + S_{\sigma_s},
\end{equation}
where $d_0$ is a free-space reference distance, $PL_0$ is the corresponding free-space loss, $n_p$ is the path-loss exponent, and $S_{\sigma_s}$ models the log-normal shadowing. We then define the large-scale channel gain as $\beta_{g,k} \triangleq 10^{-PL(r_{g,k})/10}$, where $r_{g,k}$ is the physical distance between the BS and user $k$ in group $g$.The line-of-sight (LoS) component is modeled as
\begin{equation}\label{eq:LoS}
    \hv_{g,k}^{\rm LoS} = \alpha_{g,k}\av(\theta_{g,k}),
\end{equation}
where $\theta_{g,k}\in[0,2\pi)$ denotes the angle of departure (AoD) and the array response vector is defined as
\begin{equation}
    \av(\theta) = \left[1,e^{-j\nu_{\rm c}d_{\rm c}\sin{\theta}},\dots,e^{-j\nu_{\rm c}d_{\rm c}(N-1)\sin{\theta}}\right]^{\mathsf T},
\end{equation}
with $\nu_{\rm c}$ being the wavenumber corresponding to the carrier frequency and $d_{\rm c}$ the antenna spacing. To explicitly separate the large- and small-scale effects in the LoS coefficient, we parameterize
\begin{equation} \label{eq:alpha}
    \alpha_{g,k} \triangleq \sqrt{\beta_{g,k}}\,e^{j\vartheta_{g,k}},
\end{equation}
where $\vartheta_{g,k}\in[0,2\pi)$ represents an small scalfe fading, while the non-line-of-sight (NLoS) component $\hv_{g,k}^{\rm NLoS}$ is modeled as a circularly symmetric complex Gaussian random vector whose entries are independently distributed as $\Cc\Nc(0,\beta_{g,k})$. 
It has been shown in~\cite{rappaport2013millimeter} that mmWave propagation environments are typically dominated by strong LoS components, since severe reflection losses cause the NLoS components to be attenuated by more than $10$ dB relative to the LoS component. 
Accordingly, we assume a strongly LoS-dominant regime with $\kappa\gg 1$ for the design of the proposed transmission scheme, such that $\hv_{g,k} \approx \hv_{g,k}^{\rm LoS}$.\footnote{The impact of finite Rician $\kappa$, i.e., non-negligible NLoS components, is further evaluated in the simulation results.}
Finally, to maintain full consistency with the channel model adopted throughout the main text and the simulation setup, we specialize the large-scale gain to a simple distance-dependent model without additional shadowing, i.e., $\beta_{g,k}=r_{g,k}^{-1}$, which reduces \eqref{eq:alpha} to $\alpha_{g,k}=r_{g,k}^{-1/2}e^{j\vartheta_{g,k}}$.
}

We adopt a quasi-static block fading model, where the channel vectors $\mathbf h_{g,k}$ remain constant over one coherence block consisting of $T_c$ symbol intervals (time slots) and change independently across different blocks. In conventional CIST-based downlink beamforming, each coherence block typically consists of an overhead phase, including beam training, channel estimation, and possibly feedback, followed by a payload (data) transmission phase~\cite{alkhateeb2017initial}. A precoder design based on instantaneous CSIT is then applied during the data transmission phase within the same coherence block.

\subsection{Multi-Slot Frame Structure}\label{subsec:system2}


Motivated by the highly directional nature of mmWave and sub-THz communications, multicast transmission is typically realized through a sequence of beamformed transmissions over time, rather than via a single wide-beam broadcast~\cite{chukhno2021efficient}. This {\em sequential} transmission structure is particularly relevant in overloaded regimes, where the number of users significantly exceeds the spatial degrees of freedom (DoF) at the transmitter. To capture this operational characteristic and to establish a fair time-division MGM baseline for overloaded scenarios, we adopt a time-slotted transmission model.

Specifically, we consider a transmission frame consisting of $S$ consecutive time slots, indexed by $s \in [S]$. The downlink channels are assumed to remain constant over the entire $S$-slot frame, i.e., the frame duration is smaller than the channel coherence time. This quasi-static-within-frame assumption will be explicitly exploited in the subsequent formulation of time-division MGM transmission schemes. 

In time slot $s\in[S]$, the BS simultaneously transmits $G$ multicast data streams (i.e., one stream per group) using a set of beamforming vectors
$\{\mathbf{w}^{(s)}_g\in\mathbb{C}^{N}: g\in[G]\}$.
The transmitted signal in slot $s$ is given by
\begin{equation}
\mathbf{x}^{(s)}=\sum_{g=1}^{G}\mathbf{w}^{(s)}_g s_g,
\label{eq:tx_slot}
\end{equation}
where $s_g\in\mathbb{C}$ denotes the multicast symbol intended for group $g$, normalized such that $\mathbb{E}[|s_g|^2]=1$. The received signal at user $k\in [K_g]$ in group $g$ during slot $s$ is expressed as
\begin{equation}
  y_{g,k}^{(s)} = \mathbf h_{g,k}^{\sf H}\sum_{g=1}^G\mathbf w_{g}^{(s)}s_g + n_{g,k}^{(s)},\label{eq:received signal}
\end{equation}
where $n^{(s)}_{g,k}\sim\mathcal{CN}(0,\sigma^2)$ denotes the additive receiver noise. We impose a per-slot sum transmit power constraint at the BS, given by
\begin{equation}
\sum_{g=1}^{G}\big\|\mathbf{w}^{(s)}_g\big\|^2\le P_t,\qquad \forall s\in[S].
\label{eq:power_slot}
\end{equation}

\section{Time-Division MGM Framework}\label{sec:3}

We clarify why the multi-slot transmission framework introduced in Section~\ref{sec:2} is necessary in {\em overloaded} systems through max-min fairness degrees of freedom (MMF--DoF) analysis. Specifically, we show that, in the overloaded regime of interest, the MMF--DoF of the conventional single-slot MGM baseline (i.e., operating with $S=1$) collapses to zero. We then formally formulate the corresponding time-division MGM problem and propose a novel approach to efficiently address it.

\subsection{Fundamental Limitations of Single-Slot MGM}
\label{subsec: 3-1}
We begin by considering the conventional \emph{single-slot} MGM baseline, which corresponds to the case $S=1$ in the system model. Our objective is to characterize its fundamental limitations in overloaded LoS-dominant mmWave systems, where $N\leq K=\sum_{g=1}^{G} K_g$. In this approach, all users across all multicast groups are served simultaneously within a single time slot using one set of multicast beamformers $\{\wv_g: g\in [G]\}$. 

Following the system model in Section~\ref{subsec:system2} and under single-slot operation, the received signal at user $k$ in group $g$ is given by
\begin{equation}
  y_{g,k} = \mathbf h_{g,k}^{\sf H}\sum_{g=1}^G\mathbf w_{g}s_g + n_{g,k},
\end{equation} which consists of the desired multicast signal intended for group $g$, inter-group interference from other multicast streams, and additive noise. Accordingly, the instantaneous signal-to-interference-plus-noise ratio (SINR) experienced by user $k$ in group $g$ is expressed as
\begin{equation}
  \mathrm{SINR}_{g,k}
  = \frac{|\hv_{g,k}^{\sf{H}} \wv_g|^2}{\sum_{g' \ne g} |\mathbf h_{g,k}^\mathsf{H} \mathbf w_{g'}|^2 + \sigma^2 }. \label{eq:SINR}
\end{equation}
Given the transmit power constraint $P_t$ at the BS, the single-slot MMF rate is defined as
\begin{equation}
  R_{\mathrm{MMF}}^{(1)}(P_t)\triangleq \max_{\{\mathbf w_g\}} \min_{g \in [G],\, k \in [K_g]}
  \log_2\bigl(1 + \mathrm{SINR}_{g,k}\bigr).
\end{equation} 


\begin{definition}[Single-slot MMF--DoF] To characterize the asymptotic fairness performance of the single-slot MGM baseline in the high-SNR regime, we define the MMF--DoF as
\begin{equation}
  d_{\mathrm{MMF}}^{(1)} \triangleq \lim_{P_t \to \infty}\frac{R_{\mathrm{MMF}}^{(1)}(P_t)}
  {\log_2 P_t}.\label{DoF}
\end{equation} This definition captures the high-SNR scaling behavior of the worst-user rate under single-slot MGM operation and serves as a fundamental performance metric for assessing fairness in overloaded mmWave systems. \hfill $\blacksquare$
\end{definition}
\vspace{0.1cm}

In the following, we show that the single-slot MGM baseline suffers from a fundamental limitation in overloaded mmWave systems: when the number of users significantly exceeds the number of transmit antennas, the resulting MMF--DoF collapses to zero. This result reveals that, regardless of beamformer optimization, simultaneously serving all users within a single time slot is fundamentally inefficient in the overloaded regime.

\subsubsection{Upper Bound on the Single-Slot MMF--DoF}

To characterize the best possible performance of the single-slot MGM scheme, we first consider an \emph{idealized} benchmark scenario in which all users within the same multicast group share exactly the same channel direction. Specifically, we assume that
\begin{equation}
  \mathbf h_{g,k} = \mathbf h_g, \; \forall k \in [K_g],\; \forall g \in [G],
  \label{eq:ideal_group_channel}
\end{equation}
so that each multicast group $g$ can be represented by a single rank-1 channel vector $\hv_g \in \mathbb{C}^{N}$.

Let $\Hm \triangleq [\hv_1,\dots,\hv_G] \in \mathbb{C}^{N \times G}$. As assumed in the system model, we consider the case $G \le N$, under which $\Hm$ has full column rank.
This corresponds to an ideal situation where the group channel vectors are mutually linearly independent and, in the best case, span $G$ orthogonal spatial dimensions. In this setting, it is well known that zero-forcing (ZF) beamforming can completely eliminate inter-group interference~\cite{yoo2006optimality}. One such ZF solution is given by
\begin{equation}
  \Wm_{\mathrm{ZF}}  \triangleq \Hm \bigl(\Hm^{\sf{H}}\Hm\bigr)^{-1} \Dm,
\end{equation}
where $\Dm = \mathrm{diag}(d_1,\dots,d_G)$ is a diagonal matrix containing the power-allocation coefficients. These coefficients are chosen to satisfy the transmit power constraint $\sum_{g=1}^G \|\wv_g\|^2 \le P_t$, where $\wv_g$ denotes the $g$-th column of $\Wm_{\mathrm{ZF}}$. By construction, the ZF beamformer satisfies
\begin{equation}
  \mathbf h_{g'}^{\sf{H}} \mathbf w_g = 0,
  \quad \forall g' \ne g,
\end{equation}
which implies that each multicast group experiences \emph{no} inter-group interference. 

Consequently, in the high-SNR regime, the SINR of any user in group $g$ becomes
\begin{equation}
  \mathrm{SINR}^{\mathrm{(ideal)}}_{g,k}= 
  \frac{|\mathbf h_g^{\sf{H}} \mathbf w_g|^2}{\sigma^2} \triangleq
    \frac{c_g P_t}{\sigma^2},
\end{equation}
where $c_g>0$ is an effective channel gain determined by
the group channel vector $\hv_g$ and the power-allocation matrix $\Dm$. The achievable rate of every user scales as
\begin{equation}
  \log_2\bigl(1+\mathrm{SINR}^{\mathrm{(ideal)}}_{g,k}\bigr)
  = \log_2 P_t + \mathcal{O}(1),
\end{equation}
which yields the MMF--DoF of
\begin{equation}
  d_{\mathrm{MMF}}^{(1)} = 1.
\end{equation}

This result provides a fundamental {\em upper bound} on the MMF--DoF of single-slot MGM, attainable only under the highly idealized conditions of perfect intra-group channel alignment and full-rank inter-group channel directions.



\subsubsection{Realistic LoS geometry and MMF--DoF Collapse}

In practical overloaded mmWave systems, the idealized assumptions in \eqref{eq:ideal_group_channel} are rarely satisfied. Users within the same multicast group are randomly distributed over the cell and typically experience different AoDs and propagation distances. Thus, the channel of user $k$ in group $g$ can be approximated as
\begin{equation}
  \mathbf h_{g,k} \approx \alpha_{g,k} \mathbf a(\theta_{g,k}),
\end{equation} where $\theta_{g,k}$ and $r_{g,k}$ denote the {\em user-specific} AoD and distance, respectively, and $\av(\theta)$ is the LoS array steering vector defined in Section~\ref{sec:2}. Under this realistic geometry, the users within multicast group $g$ span a higher-dimensional \emph{group subspace}
\begin{equation}
  \mathcal{V}_g \triangleq \mathrm{span}\bigl\{
    \mathbf a(\theta_{g,1}),\dots,\mathbf a(\theta_{g,K_g})
  \bigr\} \subseteq \mathbb{C}^{N}.
\end{equation}
In general, these group subspaces $\{\mathcal{V}_g\}_{g=1}^G$ are not mutually orthogonal and may significantly \emph{overlap} in the angular domain. In the overloaded systems with $K \gg N$, it is therefore fundamentally impossible to design beamformers that simultaneously align with all users in one group while remaining orthogonal to users in other groups.

To highlight the geometry structure of interference, we express the beamformers as 
\begin{equation}
    \mathbf w_{g}=\sqrt{P_t/G}\mathbf b_g,\; \|\mathbf b_g\|=1, 
\end{equation} which explicitly factors out the transmit power scaling. 
Substituting this representation into the SINR expression yields
\begin{equation}
    \mathrm {SINR}_{g,k} = \frac{\frac{P_t}{G}|\alpha_{g,k}|^2 \left|\mathbf a(\theta_{g,k})^{\mathsf T}\mathbf b_{g}\right|^2}{\frac{P_t}{G}|\alpha_{g,k}|^2\sum_{g'\neq g}\left|\mathbf a(\theta_{g,k})^{\mathsf T}\mathbf b_{g'}\right|^2 + |\sigma|^2}.\label{eq:SINR2}
\end{equation} 
In the high-SNR regime, as $P_t\to\infty$, the noise term becomes negligible and the SINR converges to the \emph{geometric SIR} (GeoSIR)
\begin{equation}
\lim_{P_t\to\infty}\mathrm {SINR}_{g,k}=\mathrm {GeoSIR}_{g,k} \triangleq \frac{\left|\mathbf a(\theta_{g,k})^{\mathsf T}\mathbf b_{g}\right|^2}{\sum_{g'\neq g}\left|\mathbf a(\theta_{g,k})^{\mathsf T}\mathbf b_{g'}\right|^2}.\label{eq:geoSIR}
\end{equation}

For generic user geometries where the group subspaces overlap, both the desired signal power and the aggregate inter-group interference scale linearly with $P_t$. Consequently, the SINR of the worst-case user remains bounded, and the MMF rate saturates to a constant. This directly leads to 
\begin{equation}
  d_{\mathrm{MMF}}^{(1)} = 0.
\end{equation}
This phenomenon is consistent with the MMF--DoF collapse previously observed in overloaded systems under i.i.d. Rayleigh fading, where user channels become asymptotically independent~\cite{joudeh2017rate}. However, in the mmWave regime, the underlying mechanism is fundamentally different: the collapse arises from the strong angular structure of LoS steering vectors and the inevitable overlap of group subspaces in the spatial domain.

\begin{figure}[t]
\centering
\includegraphics[width=0.95\linewidth]{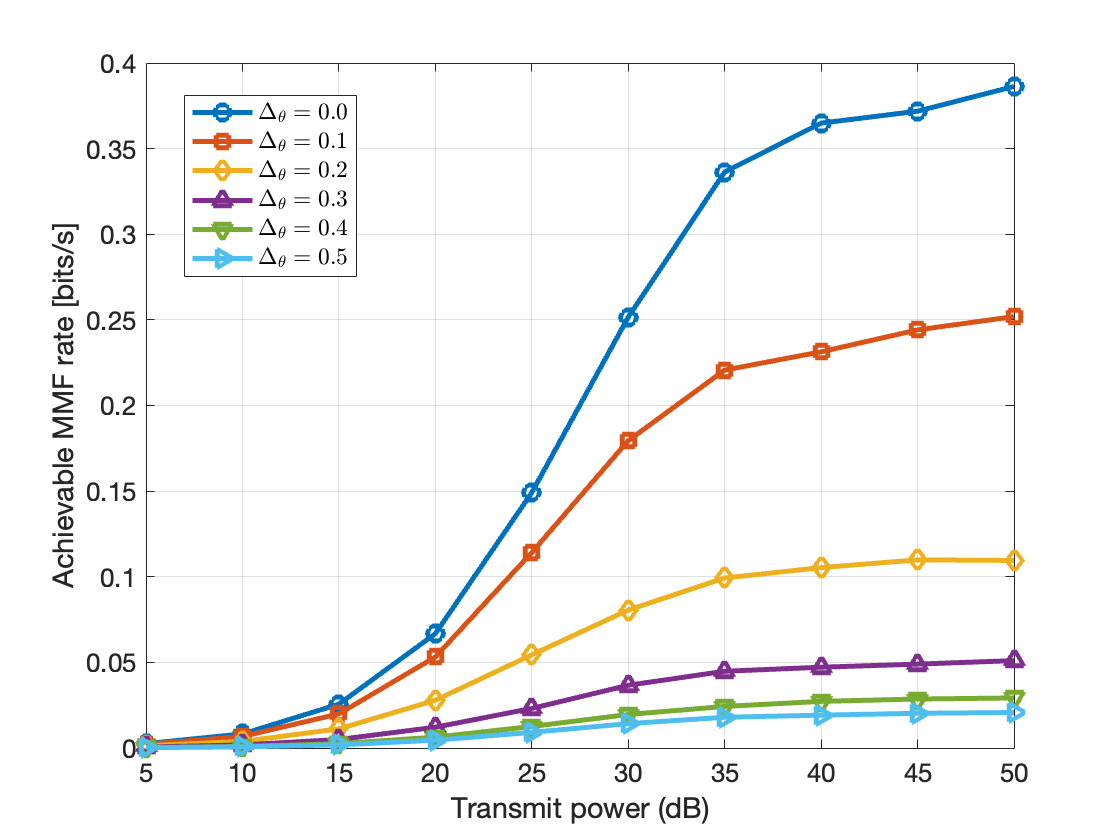}
\caption{Single-slot MGM MMF rate versus $\SNR = P_t/\sigma^2$ for different angular sector widths $\Delta_\theta \in \{0, 0.1, 0.2, 0.3, 0.4, 0.5\}$ radian. System parameters: $N=8$ transmit antennas, $G=6$ multicast groups, and $K_g=8$ users per group.
}
\label{fig:1}
\end{figure}

To verify the above MMF--DoF collapse under {\em realistic} overloaded mmWave conditions, we present a numerical example in Fig.~\ref{fig:1}. The detailed channel model and simulation settings are identical to those in Section~\ref{sec:sim}. In each realization, $G$ multicast groups are served by the BS equipped with a uniform linear array (ULA), and the AoDs of users within each group are generated around a randomly selected group-center angle. Specifically, the group-center angles are drawn independently and uniformly from $[0,2\pi)$, i.e., $\phi_g \sim \mathrm{Unif}[0,2\pi)$, and the users in group $g$ are uniformly distributed within an angular sector of width $\Delta_\theta$ centered at $\phi_g$. For transmission, the BS designs single-slot ZF beamformers based on the group-representative channels, defined as the dominant singular vectors of per-group channel matrices, and scales them to satisfy the sum-power constraint. 

Fig.~\ref{fig:1} shows the single-slot MMF rate as a function of the SNR, defined as $\SNR=P_t/\sigma^2$, for different angular spreads $\Delta_\theta$. When $\Delta_\theta = 0^\circ$, corresponding to the ideal case in which all users within each group are perfectly aligned, the worst-case user rate grows linearly with $\log_2 P_t$, which is consistent with an MMF-DoF of one (i.e., $d_{\mathrm{MMF}}^{(1)} = 1$) under ZF beamforming. In contrast, for any nonzero angular spread $\Delta_\theta > 0^\circ$, the worst-case user rate quickly saturates as the $\SNR$ increases, and the empirical MMF--DoF collapses to zero. Moreover, as the angular spread increases from $0$ to $0.5$ rad, the saturation level decreases monotonically. This behavior reflects the increased overlap among the LoS-induced group subspaces and the resulting growth of inter-group interference. 

These results confirm that, under realistic user geometries, the single-slot MGM baseline cannot achieve a non-zero MMF--DoF. This fundamental limitation directly motivates the proposed time-division MGM framework, which aims to create additional degrees of freedom in the temporal domain.

\begin{figure*}[!t]
    \centering
    \subfigure[The conventional CSIT-Based method.]{
         \includegraphics[width=0.41\linewidth]{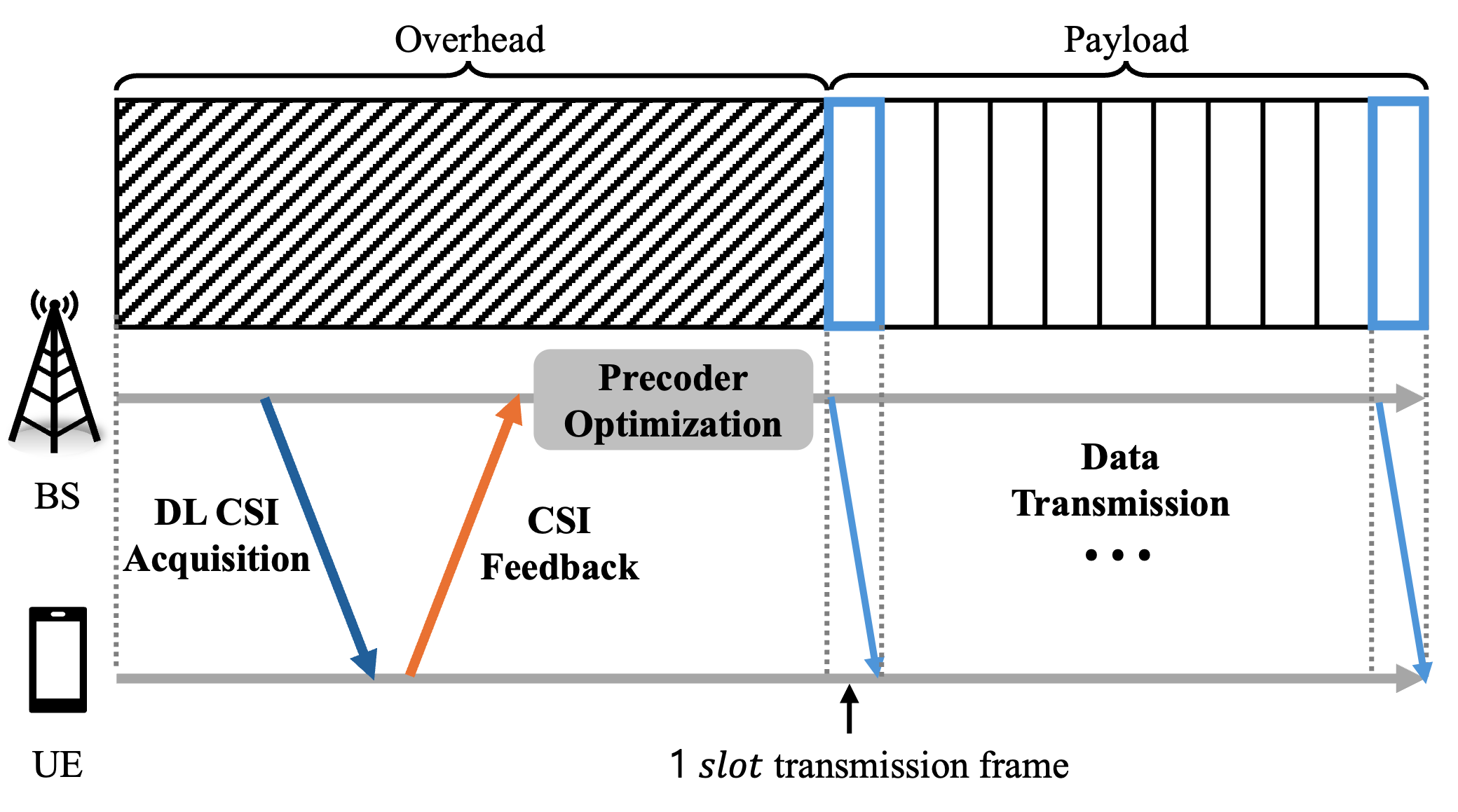}}
    \subfigure[The proposed CF-MGM method.]{\includegraphics[width=0.46\linewidth]{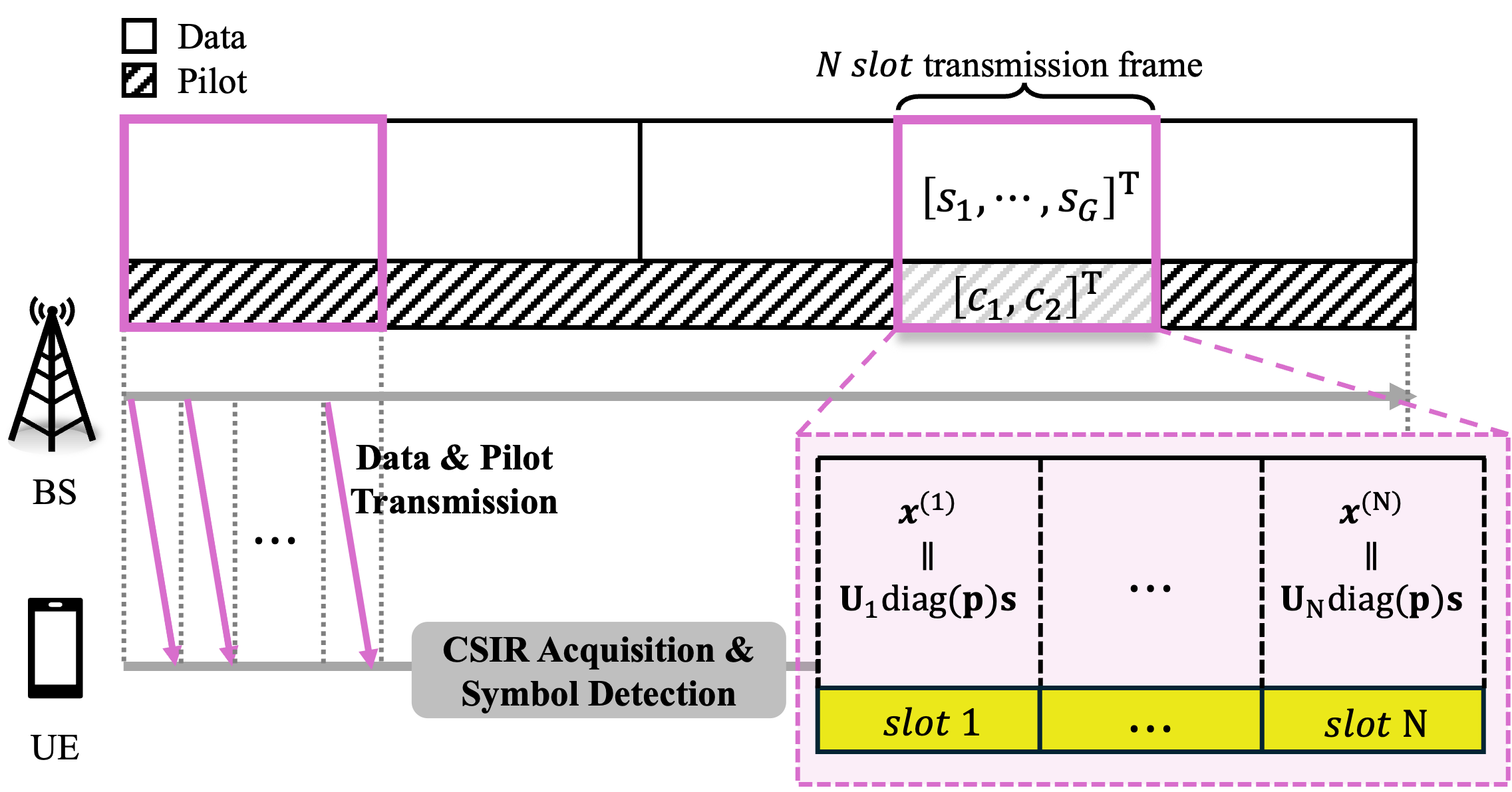}}
    \caption{Illustration of the conventional CSIT-based MGM framework, and the proposed CF-MGM framework.}
    \label{fig:simulation_graph}
\end{figure*}


\subsection{Formulation of Time-division MGM Problem}\label{subsec:3-2}
The above analysis demonstrates that, under realistic LoS user geometries, the conventional single-slot MGM baseline ($S=1$) suffers from a collapse of the MMF--DoF. Importantly, this collapse cannot be remedied by beamforming optimization, even with full and perfect CSIT, revealing an intrinsic {\em geometry-induced limitation} of single-slot operation. This observation indicates that improving fairness in overloaded mmWave systems requires more than sophisticated beamformer design alone. In particular, how users are partitioned across time slots plays a critical role. By appropriately distributing users over multiple slots, the system can exploit the underlying channel geometry to partially decouple user channels across time. Such time-division operation introduces additional design freedom, which can be leveraged to mitigate inter-group interference and improve the achievable MMF rate.

Motivated by this insight, we generalize the conventional single-slot MGM baseline to a time-division framework with $S>1$ slots per frame, as introduced in Section~\ref{sec:2}. Instead of serving all $K_g$ users of each group simultaneously, the BS schedules users across multiple slots and designs {\em slot-dependent} beamformers. This enables the system to exploit time-domain degrees of freedom through user scheduling and frame-level averaging, which is particularly crucial in overloaded mmWave MGM scenarios where satisfying the worst-case user performance  within a single slot is overly restrictive.


Building on the time-slotted model introduced in Section~\ref{sec:2}, for each slot $s\in[S]$, the transmitted signal $\xv^{(s)}$ and the per-slot sum-power constraint are given by \eqref{eq:tx_slot} and \eqref{eq:power_slot}, respectively. Likewise, the received signal and the instantaneous SINR of user $k$ in group $g$ during slot $s$ take the same form as in the single-slot case in~\eqref{eq:SINR};  we denote the resulting SINR by $\mathrm{SINR}^{(s)}_{g,k}$.

To model slot-wise user activation, we introduce binary scheduling variables $z_{g,k}^{(s)}\in\{0,1\}$, where $z_{g,k}^{(s)}=1$ indicates that user $k$ in group $g$ is scheduled in slot $s$. The frame-averaged rate of user $k$ in group $g$ is then defined as
\begin{equation}
\bar r_{g,k}\triangleq \frac{1}{S}\sum_{s=1}^{S} z_{g,k}^{(s)}\log_2\!\left(1+\mathrm{SINR}^{(s)}_{g,k}\right).
\label{eq:td_rate}
\end{equation}
With these definitions, the {\em time-division MGM MMF problem} can be formulated as the following max--min optimization:
\begin{equation}
  \label{prob:TD-MGM}
  \begin{aligned}
   \max_{t,\mathbf w_g^{(s)},z_{g,k}^{(s)}} t \\
  \text{subject to}&\;
    \bar r_{g,k}\ge t,
    \; \forall g\in[G],\; \forall k\in[K_g],\\
  & \sum_{g=1}^{G} \bigl\|\mathbf w_g^{(s)}\bigr\|^2
    \le P_t,\;
    \forall s\in[S], \\
  & z_{g,k}^{(s)} \in \{0,1\},\;
     \forall g\in[G],\; \forall k\in[K_g],\; \forall s\in[S]. 
\end{aligned}
\end{equation}
Problem~\eqref{prob:TD-MGM} precisely captures the MMF objective of a CSIT-based time-division MGM system. However, it is highly challenging to solve in practice. The problem is a {\em mixed-integer nonconvex program}, in which the binary scheduling variables $z_{g,k}^{(s)}$, the slot-dependent multicast beamformers$\{\wv_g^{(s)}\}$, and the underlying LoS geometry are strongly coupled.

\subsection{Proposed CSIT-free Approach}\label{subsec:idea}

To overcome the above difficulties, we adopt a practically motivated and computationally efficient approach to MMF optimization. Instead of explicitly optimizing the binary scheduling variables $z_{g,k}^{(s)}$ jointly with the beamformers—which leads to a mixed-integer nonconvex problem typically requiring iterative optimization—we propose a {\em CSIT-free} multi-group multicast framework, termed CF–MGM, that provides an efficient and scalable solution to the MMF objective. 

The key idea of CF--MGM is to bypass explicit user scheduling and instantaneous beamformer adaptation altogether. In the proposed framework, all users are conceptually active in every slot, and inter-group interference is eliminated {\em by construction} through a carefully designed $N$-slot transmission structure with CSIT-free beamforming design. Thus, the BS does not require any instantaneous channel state information, which makes the proposed approach particularly attractive for overloaded mmWave systems where CSIT acquisition incurs substantial overhead. This structural interference cancellation fundamentally simplifies the original time-division MMF problem. Specifically, the MMF objective can be expressed directly in terms of group-level transmit power variables, without involving beamforming or scheduling decisions. Consequently, the original mixed-integer nonconvex formulation reduces to a low-dimensional max–-min power-allocation problem with a convex feasible set, which admits an optimal closed-form solution.

While CF--MGM does not aim to achieve the global optimum of the original time-division MMF problem, it provides a highly efficient solution that captures the essential fairness objective. In sharp contrast to the conventional single-slot MGM baseline, whose MMF--DoF collapses to zero under realistic LoS geometries, the proposed framework guarantees a \emph{strictly positive} MMF--DoF in overloaded mmWave systems. Extensive simulation results in Section~\ref{sec:sim} further demonstrate that CF--MGM achieves substantial performance gains over existing baselines, thereby validating its effectiveness as a practical and theoretically grounded solution to MMF optimization.

\begin{remark}\label{remark:DoF-Collapse}
    By adopting a time-division (or multi-slot) MGM approach, the conventional CSIT-based schemes, in principle, avoid the MMF--DoF collapse, provided that the number of slots per frame is sufficiently large such that $NS \geq K=\sum_{g=1}^{G} K_g$. For example, in an overloaded mmWave setting with $N=8$, $G=6$, and $K_g=24$ for all $g \in [G]$, avoiding DoF collapse requires $S=18$ time slots. While this approach can theoretically restore a nonzero DoF, it comes at a substantial cost: excessive time sharing (i.e., a relatively large number of slots $S$ in overloaded systems) causes a severe loss in throughput. Therefore, such a brute-force time-division solution is fundamentally ill-suited for overloaded mmWave systems, where both spectral efficiency and latency are critical. These limitations motivate the need for a fundamentally different transmission paradigm, which we address by introducing the proposed CSIT-free MGM framework.
\end{remark}


\section{Proposed CF-MGM Framework}\label{sec:4}

We describe the proposed CF--MGM, introduced in Section~\ref{subsec:idea}, which provides a structured and scalable solution to the time-division MGM problem in \eqref{prob:TD-MGM}. Unlike conventional approaches that rely on CSIT-dependent beamformer design and slot-wise user scheduling, CF--MGM adopts a {\em deterministic multi-slot precoding structure} within each transmission frame. A key feature of CF--MGM is that it completely eliminates the need for per-block instantaneous CSIT. The BS employs a fixed frame-level transmission structure, thereby avoiding per-block beam training, channel estimation, and feedback. As a result, a large fraction of the coherence block can be devoted to downlink data transmission, which is particularly beneficial in overloaded mmWave systems with short coherence times and high training overhead. At the receiver side, only lightweight channel state information at the receiver (CSIR) is required. This is obtained via a small number of embedded downlink pilots and simple linear combining across slots, without any CSIT feedback. Therefore, CF--MGM substantially reduces signaling overhead while maintaining low receiver complexity. 

{In the proposed method, we set $S=N$ and, for ease of exposition, focus on the case $N=G+2$. When $N \gg G$ (e.g., when the number of transmit antennas significantly exceeds the number of groups), the framework naturally extends by transmitting multiple symbols per group within a frame. the same transmission structure naturally supports multiple multicast symbols per group within a single frame, since the $N$-dimensional symbol vector can embed $N-2$ data symbols with a fixed 2 pilot symbols. Consequently, the frame length does not imply a loss in spectral efficiency in the large $N$ regime.} For clarity, Fig.~\ref{fig:simulation_graph} illustrates the basic transmission unit of the conventional CIST-based MGM in comparison with the proposed CF--MGM framework. CF--MGM adopts an $N$-slot transmission frame as the basic transmission unit. Within each frame, the BS transmits the $G$ group symbols $\{s_1,\ldots,s_G\}$ together with two pilot symbols $\{c_1,c_2\}$ for CSIR acquisition. Specifically, $N=G+2$ symbols, collected $\sv=[s_1,\ldots,s_G,c_1,c_2]^{\mathsf T}$, are conveyed over $N$ consecutive slots using deterministic multi-slot precoding. Unlike the CSIT-based baseline, where delivers $\{s_g\}$ on a per-slot basis, CF--MGM delivers them jointly over the entire frame.

In the following, we detail the transmission and combining structure of CF--MGM and show that inter-group interference is eliminated \emph{by construction}. This property reduces the original time-division MMF problem to a low-dimensional power-allocation problem with a convex feasible set, which admits a closed-form solution.


\subsection{CSIT-free Beamforming}\label{sec:4-1}

In the proposed CF-MGM framework, the BS sequentially transmits $N=G+2$ downlink signals to all users over $N$ consecutive time slots. More general choices of $N$ are possible, e.g., $N\ge G+q$, where $q\geq2$ denotes the number of pilot symbols. For each slot $n \in [N]$, the transmit signal $\xv^{(n)}$ is deterministically constructed in a CSIT-free manner as
\begin{equation}
\mathbf x^{(n)} = \mathbf U_n\mathrm{diag}(\mathbf p)\mathbf s, \label{eq:cfmgm_tx}
\end{equation}
where $\pv=[\sqrt{p_1},\dots,\sqrt{p_N}]\in \mathbb C^N$ denotes the power-allocation vector, with $p_n$ representing the power assigned to the $n$-th symbol. The {\em deterministic} precoding matrix $\Um_n$ corresponds to the $n$-th circulant permutation discrete Fourier transform ({\bf CP--DFT}) matrix, as defined in~\cite{lee2025blind}. These matrices satisfy specific orthogonality properties, which will be exploited to achieve inter-group interference cancellation without relying on instantaneous CSIT. The symbol vector $\mathbf s\in\mathbb C^N$is defined as
\begin{equation}
    \sv=[s_1,s_2,\dots,s_G,c_1,c_2]^{\mathsf T},
\end{equation} where the first $G$ entries correspond to the multicast data symbols $\{s_g : g\in[G]\}$, and the last two entries $\{c_1,c_2\}$ denote pilot symbols used for CSIR acquisition. Hence, the pilot overhead is fixed to two time slots per $N$-slot frame, corresponding to a constant fraction of $2/N$, which does not scale with the number of users. The symbol vector $\sv$ is assumed to be normalized such that $\mathbb E[\mathbf s\mathbf s^\mathsf H]=\mathrm I_N$.

Importantly, the BS does not require any instantaneous CSIT to construct the transmit signal in ~\eqref{eq:cfmgm_tx}. The matrices $\{\Um_n\}$ are deterministic and known a priori, while the power-allocation vector $\pv$ is optimized solely based on large-scale channel statistics, as will be detailed in the subsequent sections.
Under the quasi-static channel assumption over the $N$ downlink transmissions, the channel remains constant across slots, i.e., 
\begin{equation}
    \hv_{g,k}^{(s)}=\hv_{g,k}, \forall s \in [S].
\end{equation} Accordingly, the signal received by user $k$ in group $g$ during slot $n$ can be expressed as
\begin{equation}
y_{g,k}^{(n)} = \mathbf h^\mathsf T_{g,k}\mathbf x^{(n)} +z_{g,k}^{(n)}
\end{equation} where $z_{g,k}^{(n)}\sim \Cc\Nc(0,\sigma^2)$ denotes the additive noise at the receiver. 
 
By stacking the $N$ received signals across slots, we define
\begin{equation}
y_{g,k}=\left[y_{g,k}^{(1)},y_{g,k}^{(2)},\dots,y_{g,k}^{(N)}\right]^{\mathsf T}.
\end{equation} Using the proposed CF--MGM precoding in \eqref{eq:cfmgm_tx}, the stacked received signal can be rewritten as
\begin{equation}
    \mathbf y_{g,k}\stackrel{(a)}{=}\mathbf F_{g}^{\mathsf T}{\mathbf h_{g,k}} s_g\sqrt{p_g}+\sum_{g'\neq g}^N\mathbf F_{g'}^{\mathsf T}{\mathbf h_{g,k}}s(g')\sqrt{p_g'}+\mathbf z_{g,k},\label{eq:received}
\end{equation} where $\zv_{g,k}=\left[z_{g,k}^{(1)},z_{g,k}^{(2)},\dots,z_{g,k}^{(N)}\right]^{\mathsf T}$, and (a) follows from the proposed deterministic multi-slot precoding in \eqref{eq:cfmgm_tx}. Here, we define
\begin{equation}
    \mathbf F_g\triangleq\begin{bmatrix}
        \mathbf U_1(:,g)&\mathbf U_2(:,g)&\cdots&\mathbf U_N(:,g)
    \end{bmatrix}, \label{F_g}
\end{equation} which is referred to as the combining matrix for group $g$.

Each user then applies linear combining to $\yv_{g,k}$ using the estimated channel $\hat{\mathbf h}_{g,k}$ and the group combining matrix $\mathbf F_g$, yielding 
\begin{align}
&\tilde{\hv}_{g,k}^{\mathsf T}\Fm_g^{*}\yv_{g,k} = \tilde{\hv}_{g,k}^{\mathsf T}\Fm_g^{*}\Fm_{g}^{\mathsf T}{\hv_{g,k}} s_g\sqrt{p_g} \nonumber\\
&+ \sum_{g'\neq g}^N\tilde{\hv}_{g,k}^{\mathsf T}\Fm_g^{*}\Fm_{g'}^{\mathsf T}{\hv_{g,k}}\sv(g')\sqrt{p_{g'}} + \tilde{\hv}_{g,k}^{\mathsf T}\Fm_g^*\zv_{g,k}\label{eq:ybar}
\end{align}
where $\tilde{\hv}_{g,k}\eqdef\mathbf 1_{N}\oslash(\hat{\hv}_{g,k})\in\CC^{N}$ denotes the channel equalization vector, with $\oslash$ representing element-wise division.
Under the assumption of perfect CSIR,\footnote{The CSIR acquisition procedure is detailed in~\cite{lee2025blind}, where the large-scale channel gain and the AoD in \eqref{eq:LoS} are estimated using the two pilot symbols $\{c_1,c_2\}$.} it follows from the orthogonality properties of the CP-DFT structure~\cite{lee2025blind} that, for all $g'\neq g\in[N]$,
\begin{equation}
    \tilde{\hv}_{g,k}^{\mathsf T}\Fm_g^{*}\Fm_{g}^{\mathsf T}{\hv_{g,k}} = N,\mbox{and }\tilde{\hv}_{g,k}^{\mathsf T}\Fm_g^{*}\Fm_{g'}^{\mathsf T}{\hv_{g,k}} = 0.\label{eq:orthogonality}
\end{equation} 
Substituting \eqref{eq:orthogonality} into \eqref{eq:ybar},the combined signal simplifies to
\begin{equation}
\tilde{\hv}_{g,k}^{\mathsf T}\Fm_g^{*}\yv_{g,k} =  N\sqrt{p_g}s_g +\tilde{\hv}_{g,k}^{\mathsf T}\Fm_g^*\zv_{g,k}. \label{eq:ybar2}
\end{equation}

Notably, the interference from all other multicast groups is completely eliminated, while the desired multicast stream $s_g$ attains a full combining gain of $N$. Importantly, both CSIR acquisition and symbol detection are performed locally at the receiver and do not require any CSIT feedback. From the combined signal in~\eqref{eq:ybar2}, the resulting SINR under the proposed CF-MGM framework is given by
\begin{equation}
\label{eq:SINR_CFMGM}
    \mbox{SINR}_{g,k}^{\rm CF\textit{-}MGM} = \frac{\EE\left[\left|N\sqrt{ p_g}s_g\right|^2\right]}{\EE\left[\left|\tilde{\hv}_{g,k}^{\mathsf T}\Fm_g^*\zv_{g,k}\right|^2\right]}\stackrel{(a)}{\approx}\frac{|\alpha_{g,k}|N p_g}{\sigma^2},
\end{equation} 
where the approximation (a) follows from the facts that $\Fm_g^*\zv_{g,k}\sim\mathcal{CN}(\mathbf 0,\sigma^2 \Id_N)$ and $\|\tilde{\hv}_{g,k}\|_2^{2} \approx N/|\alpha_{g,k}|$ under a LoS dominant channel model (i.e., $\kappa \gg 1$). Based on the above expression, frame-averaged rate of user $k$ in group $g$, defined in~\eqref{eq:td_rate}, simplifies to
\begin{equation}
\label{rate}
  \bar r_{g,k} \triangleq 
  \log_2\!\bigl(1+\mathrm{SINR}_{g,k}^{\mathrm{CF-MGM}}\bigr).
\end{equation}
Since inter-group interference is completely canceled by the CSIT-free precoding in~\eqref{eq:cfmgm_tx}, the SINR in~\eqref{eq:SINR_CFMGM} depends solely on the allocated group power $p_g$, the large-scale channel gain $|\alpha_{g,k}|$, and the noise power $\sigma^2$. We assume that the BS has access to the slow-varying large-scale channel gains $|\alpha_{g,k}|$ (e.g., via random access-based measurements~\cite{fengler2021non,jeong2015random}), while the instantaneous small-scale channel phases remain unknown.

\subsection{Power Allocation for CF-MGM}\label{sec:4-2}

The above derivations establish a key structural property of CF--MGM: even without instantaneous CSIT, the proposed $N$-slot frame together with the receiver-side processing enables each UE to suppress the inter-group interference and reliably recover its desired multicast symbol. Importantly, once this CSIT-free decodability is guaranteed, the remaining precoding design task at the frame level reduces to determining how the total transmit power should be distributed across multicast groups (and equivalently, across symbols) so as to optimize the target system utility. 
Since the effective post-combining SINR of each group depends on its allocated transmit power and the corresponding large-scale channel gains, power allocation can be optimized using only slow-varying path-loss information available at the BS. This observation not only enables a fully CSIT-free design but also leads to a substantial simplification of the original MMF optimization problem. In the following, we develop an efficient CSIT-free power allocation rule for CF–MGM under the considered utility criterion.


We now revisit the MMF formulation in the time-division MGM framework. Recall that in optimization problem~\eqref{prob:TD-MGM}, the BS jointly optimizes the slot-wise multicast beamformers $\mathbf w_g^{(n)}$and the binary scheduling variables $z_{g,k}^{(n)}$ in order to maximize the worst frame-averaged user rate $\bar r_{g,k}$ under a per-slot power constraint. Under the proposed CF-MGM transmission structure, however, all users are effectively active in every slot and inter-group interference is completely removed. Hence, the MMF objective in~\eqref{prob:TD-MGM} can be expressed solely in terms of group-wise power allocation variables.
Letting $\mathrm {SINR}_{g,k}^{\mathrm {CF\mbox{-}MGM}}\approx a_{g,k}p_g$, the frame-averaged rate of user $k$ becomes  
\begin{equation}
  \bar r_{g,k} \triangleq 
  \log_2\bigl(1+a_{g,k}p_g\bigr).
\end{equation} 
Since $\log_2(\cdot)$ is a strictly increasing function, maximizing the worst-case user rate is equivalent to maximizing the worst-case user SINR. Consequently, the optimization problem in~\eqref{prob:TD-MGM} reduces to the following MMF power allocation problem:
\begin{equation}
\label{prob:cfmgm}
\begin{aligned}
\max_{t,\{p_g: g \in [G]\}}&\;  t \\
\text{subject to}&\; A_g{p_g}\geq t,\; \forall g\in [G],\\
& \sum_{g=1}^{G} p_g\leq P_t,\; p_g\geq 0,
\end{aligned}
\end{equation}
where the worst-case user gain in group $g$ is defined as 
\begin{equation}
    A_g\triangleq \min_{k\in [K_g]} a_{g,k}.
\end{equation}
For a fixed target value $t$ in~\eqref{prob:cfmgm}, the total transmit power required to satisfy the constraints $A_g p_g\ge t$ for all groups is given by
\begin{equation}
    \sum_{g=1}^{G} {p_g}\geq \sum_{g=1}^{G} {t}/{A_g}. 
\end{equation} 
From the power-budget constraint, any feasible $t$ must therefore satisfy  
\begin{equation}
    t\leq \frac{P_t}{\sum_{g} 1/A_g}.
\end{equation} Since the goal is to maximize $t$, the largest feasible value is obtain in closed-form by
\begin{equation}
\label{t_star}
    t^\star = \frac{P_t}{\sum_{g=1}^{G} {1}/{A_g}}.
\end{equation} 
Accordingly, the optimal power allocation for each group is 
\begin{equation}
    p_g^\star=\frac{t^\star}{A_g}.
\end{equation}
This closed-form power allocation equalizes the effective SINR across all multicast groups. The resulting achievable MMF rate, defined in~\eqref{rate}, is given by
\begin{equation}
\label{eq:MMFrate}
\min_{g\in [G],k\in[K_g]} \bar r_{g,k} =\log_2(1+t^\star).
\end{equation}

\begin{remark} 
    The proposed power allocation strategy requires only large-scale channel gains (e.g., path loss) at the BS. Such information can be readily obtained from uplink signals during the initial access phase (e.g., random access) when the users request to join a multicast stream, leveraging TDD channel reciprocity. In contrast to instantaneous CSI, these large-scale gains are quasi-static and vary over significantly longer time scales, since they are mainly governed by long-term factors such as the distance between the user and the BS. Consequently, they can be updated via infrequent user reports, imposing only negligible signaling overhead.\hfill$\blacksquare$
\end{remark}

\subsection{DoF Analysis of Proposed CF-MGM}\label{sec:4-3}

We now analyze the MMF-DoF achieved by the proposed CF-MGM framework and contrast it with the single-slot MGM baseline. Under CF-MGM, the MMF rate achieved over one $N$-slot transmission frame is given in~\eqref{eq:MMFrate}. Note that this rate is measured per frame consisting of $N$ consecutive channel uses.

To be consistent with the MMF-DoF definition~\eqref{DoF}, which normalizes the achievable rate by the number of channel uses, MMF-DoF of CF--MGM is obtained as
\begin{equation}
\begin{aligned}
  d_{\mathrm{MMF}}^{\mathrm{CF\mbox{-}MGM}}
  &=
  \lim_{P_t\to\infty}\frac{\log_2(1 + c P_t)}{N\log_2 P_t}= \frac{1}{N},
  \label{eq:dof_cfmgm}
\end{aligned}
\end{equation}
where the constant $c\triangleq \bigl(\sum_{g=1}^G1/A_g\bigr)^{-1}$ depends only on the large-scale channel gains and is independent of the transmit power $P_t$.

This result shows that the proposed CF--MGM scheme achieves a strictly positive MMF--DoF of $1/N$. In contrast, the single-slot MGM baseline analyzed in Section~III-A suffers from a complete MMF--DoF collapse, i.e., $d_{\mathrm{MMF}}^{(1)}=0$, under
realistic overloaded LoS geometries. Therefore, the structured $N$-slot CP--DFT precoding together with group-wise combining
in CF--MGM effectively prevents the DoF collapse phenomenon and enables nonzero fairness guarantees in overloaded mmWave systems.

\subsection{Extension to Weighted Power Mean Utility Functions}\label{sec:4-4}

In the previous subsections, we focused on the MMF design. From a broader perspective, we note that MMF is a special case of a \emph{general utility function} (GUF) applied to the vector of group rates. Motivated by the need to flexibly balance throughput and fairness across heterogeneous multicast groups~\cite{christopoulos2014weighted}, we consider GUF formulations based on the weighted power mean (WPM), including weighted sum rate (WSR), the weighted geometric mean (WGM), and the weighted harmonic mean (WHM). 
Thanks to the CF--MGM structure, where inter-group interference is eliminated \emph{by construction}, all these WPM-based objectives reduce to low-dimensional power allocation depending only on large-scale channel information, while preserving the same CSIT-free transmission and receiver-side combining architecture.



As shown in Section~\ref{sec:4-2}, the achievable rate for group $g$ can be expressed as
\begin{equation}
   r_g(p_g)
  =
  \min_{k\in\mathcal{K}_g}
    \log_2\!\big(1 + a_{g,k} p_g\big)
  =
  \log_2\!\big(1 + A_g p_g\big),
  \label{eq:cf_group_rate}
\end{equation}
and we collect the group rates into the vector ${\mathbf r}_g\triangleq [r_1,\dots, r_G]^\mathsf T$.  For any utility function that is monotonically increasing in each $r_g$, the underlying CF--MGM beamforming and combining structures remain unchanged. In this case, only the scalar power-allocation vector $\mathbf p \triangleq [p_1,\dots,p_G]^\mathsf T$ needs to be optimized over the power simplex
\begin{equation}
  \mathcal{P}
  \triangleq
  \left\{
    \pv \in \mathbb{R}_+^G \,:\,
    \sum_{g=1}^G p_g \le P_t
  \right\}.
  \label{eq:power_simplex}
\end{equation} To capture different fairness--throughput tradeoffs within a unified framework, we adopt a WPM-based GUF~\cite{fang2025optimal}, defined as
\begin{equation}
  U_q(\rv)
  \triangleq
  \begin{cases}
    \Big(
      \sum_{g=1}^G \zeta_g r_g^q
    \Big)^{1/q},
      & q \neq 0,
      \\
    \Big(
      \Pi_{g=1}^G r_g^{\zeta_g}
    \Big)^{1/G},
      & q = 0,
  \end{cases}
  \label{eq:wpm_def}
\end{equation}
where $\zeta_g > 0$ denotes the weight of group $g$. 
The parameter $q$ controls the throughput--fairness tradeoff and several several important special cases include: $q=1$ (WSR), $q=0$ (WGM), $q=-1$ (WHM), and $q\to-\infty$ (MMF), i.e., $U_q(\rv)\to \min_g r_g$.
Accordingly, the CF-MGM power-allocation problem can be written in the generic form:
\begin{align}
  \max_{\pv \in \mathcal{P}}~
    & U_q\big( r_1(p_1),\dots,r_G(p_G) \big).
    \label{eq:wpm_generic_prob}
\end{align}
For the WPM-based utilities, \eqref{eq:wpm_generic_prob} is a convex optimization problem in $\pv$ with affine constraints; hence, the KKT conditions are sufficient for global optimality. In particular, introducing the dual variable $\lambda\ge 0$ for the sum-power constraint, the stationarity condition yields a per-group relation between $p_g^\star$ and $\lambda^\star$, while $\lambda^\star$ is uniquely determined such that $\sum_{g=1}^G p_g^\star = P_t$. As a representative example, for $q=1$ (WSR), stationarity gives
\begin{equation}
\frac{\zeta_g A_g}{(1+A_g p_g^\star)\ln 2}=\lambda^\star,
\end{equation}
and $\lambda^\star$ is chosen such that $\sum_g p_g^\star=P_t$.
For $q=0$ (WGM) and $q=-1$ (WHM), the same KKT-based approach applies: the stationarity condition yields a per-group closed-form relation $p_g^\star(\lambda)$ for each group, and $\lambda^\star$ can be efficiently obtained via a one-dimensional bisection search to satisfy $\sum_{g=1}^G p_g^\star(\lambda^\star)=P_t$.

\begin{figure}[t]
\centering
\includegraphics[width=0.95\linewidth]{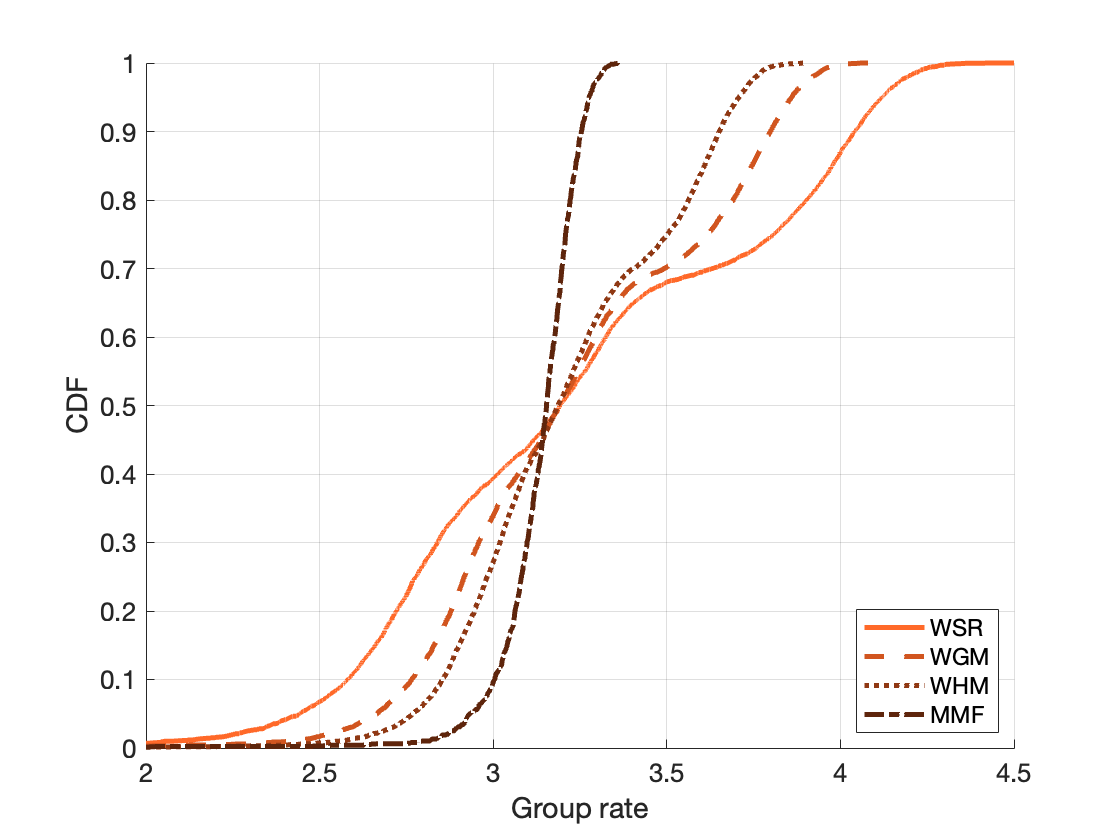}
\caption{Empirical CDF of group rates under different utility functions.}
\label{group_CDF}
\end{figure}

Fig.~\ref{group_CDF} plots the empirical cumulative distribution function (CDF) of the per-group rate $r_g$ under four representative utility choices (WSR, WGM, WHM, and MMF). To expose the impact of different fairness objectives, we form three distance clusters with two multicast groups in each range $[30,50]$m, $[60,80]$m, and $[90,110]$m, which creates a mixture of ``strong'' and ``weak'' groups in each Monte Carlo realization. Under this setting, WSR tends to favor stronger groups (heavier upper tail), whereas WGM/WHM increasingly penalize low-rate groups (right-shifting the lower tail), thereby providing a flexible mechanism to balance throughput and fairness within the same CSIT-free CF--MGM transmission architecture.
\section{Simulation Results}\label{sec:sim}

In this section, we evaluate the performance of the proposed CF-MGM framework through comparisons with several benchmark schemes.  We consider a single-cell downlink MGM system with $N=8$ transmit antennas and $G=6$ multicast groups. Unless otherwise specified, each group contains $K_g=24$ users, the Rician factor is set to $\kappa=12$, the receiver noise variance is $\sigma^{2}=1$, and the transmit SNR at the BS is fixed to $P_t/\sigma^2=30$ dB. {All simulations were conducted using MATLAB R2024b on macOS with an Apple M4 chip and 16 GB of memory; CVX-based problems were solved using SDPT3.}

As discussed in Section~\ref{subsec:3-2}, all benchmark schemes are implemented under a {\em time-division} framework to accommodate the {\em overloaded} user regime. Specifically, each scheme operates over $N=8$ time slots per transmission frame, and in each slot every multicast group serves $K_g/N = 3$ users, so that all $K_g$ users are scheduled exactly once per frame.
Unless otherwise stated, all benchmark schemes follow this $N$-slot time-division operation, thereby enabling fair and consistent performance comparisons. {It is worth emphasizing that this multi-slot operation is introduced solely to reduce the excessive instantaneous user loading in overloaded scenarios. Nevertheless, the conventional CSIT-based benchmark schemes still operate in the MMF–DoF collapse regime. This is because, in each slot, the total number of simultaneously served users across all groups remains $G(K_g/N)$, which exceeds the avaliable spatial DoF whenever $N < G(K_g/N)$. Consequently, even with time-division scheduling, inter-group interference cannot be fully eliminated by beamforming alone, and the worst-case user rate continues to saturate at high SNR.
As noted in Remark~\ref{remark:DoF-Collapse}, completely avoiding the DoF collapse would require a much larger number of time slots, which drastically reduces the number of users served per slot and incurs severe throughput loss due to excessive time sharing. In contrast, the proposed CF--MGM framework can simultaneously serve multiple users per group (e.g., three users per group in our simulations) while still avoiding MMF–DoF collapse, owing to its structured multi-slot transmission and CSIT-free interference elimination. This fundamental distinction enables CF--MGM to achieve both scalable fairness and significantly higher spectral efficiency in overloaded mmWave systems.
}

Accordingly, both the proposed CF-MGM and the benchmark schemes employ identical frame structures and the same per-slot transmit power $P_t$, ensuring fair and meaningful performance comparisons. For the proposed CF--MGM, the achievable MMF rate is computed according to the formulation in \eqref{rate}, while for the benchmark schemes it is obtained as the frame-averaged MMF rate over the $N$ slots. User locations are generated independently for each realization. Specifically, for every group $g \in [G]$ and user $k \in [K]$, the AoD is drawn uniformly as $\theta_{g,k} \sim [0,2\pi]$, and the user distance is uniformly distributed as $r_{g,k} \sim [50,100]$\,m. All benchmark schemes considered in the simulations are summarized as follows : 
\begin{itemize}
\item \textbf{SDR-GR~\cite{chang2008approximation}:} One semidefinite relaxation (SDP) is solved, followed by eigen-decomposition to extract randomized beam candidates, where a LP with bisection is executed to determine $t$ for each candidate. 
\item \textbf{RS-WMSE~\cite{joudeh2017rate}:} A rate-splitting-based approach in which the MMF problem is reformulated into a weighted mean-square error (WMSE) minimization and solved iteratively via alternating optimization.
\item \textbf{SCA-HFPI~\cite{fang2025optimal}:} This method employs the successive convex approximation (SCA) outer loop as the standard SCA in \cite{zhou2020intelligent}, but each inner subproblem is solved via hyperplane fixed-point iteration (HFPI) using closed-form updates without CVX solvers. 
\item \textbf{Standard SCA~\cite{zhou2020intelligent}:} A conventional SCA-based algorithm in which the nonconvex MMF problem is approximated by a sequence of convex subproblems, each solved using a generic convex optimization solver.
\item \textbf{PSA~\cite{zhang2022fast}:} A projected subgradient method that directly updates the multicast beamformers in the primal domain based on the worst-user SINR subgradient, followed by projection onto the total power constraint.
\end{itemize} 
Finally, we evaluate the expected achievable MMF rate of all schemes, where the expectation is computed via Monte Carlo simulations with $10^3$ trials.

\begin{figure}[t]
\centering
\includegraphics[width=0.95\linewidth]{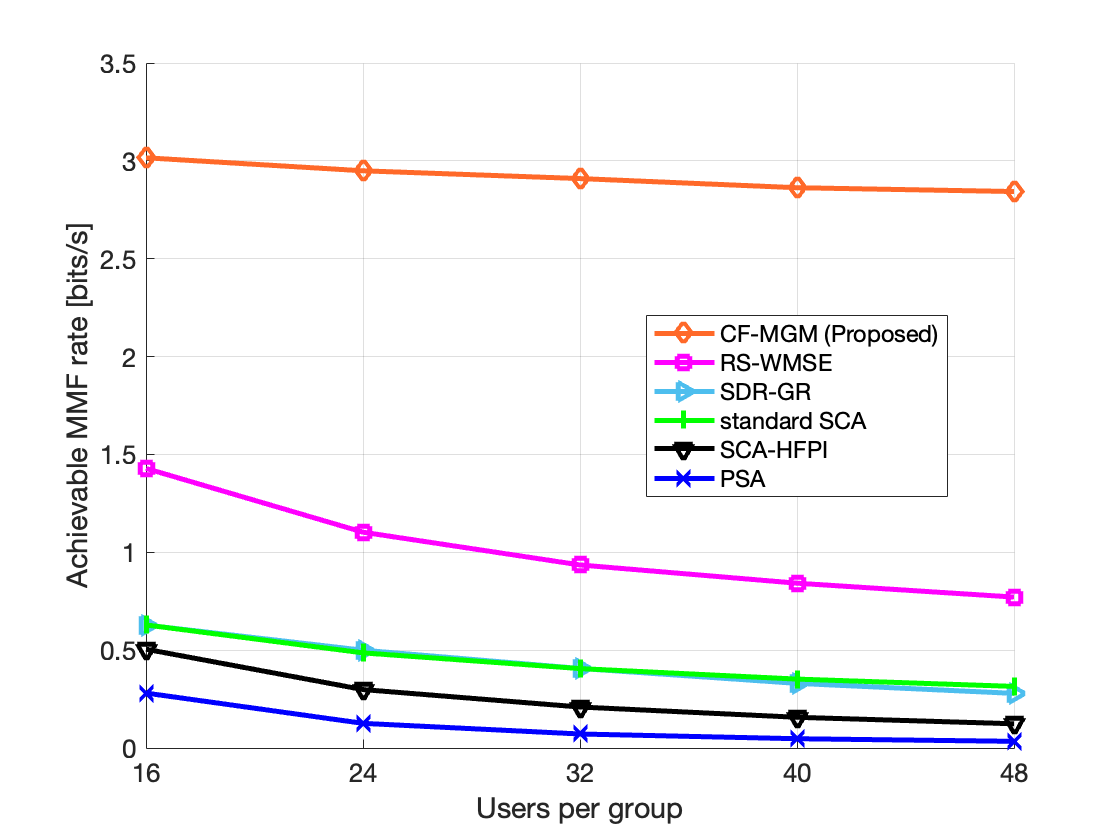}
\caption{Minimum rate versus number of users per group.}\label{UpG}
\end{figure}

\begin{figure}[t]
\centering
\includegraphics[width=0.95\linewidth]{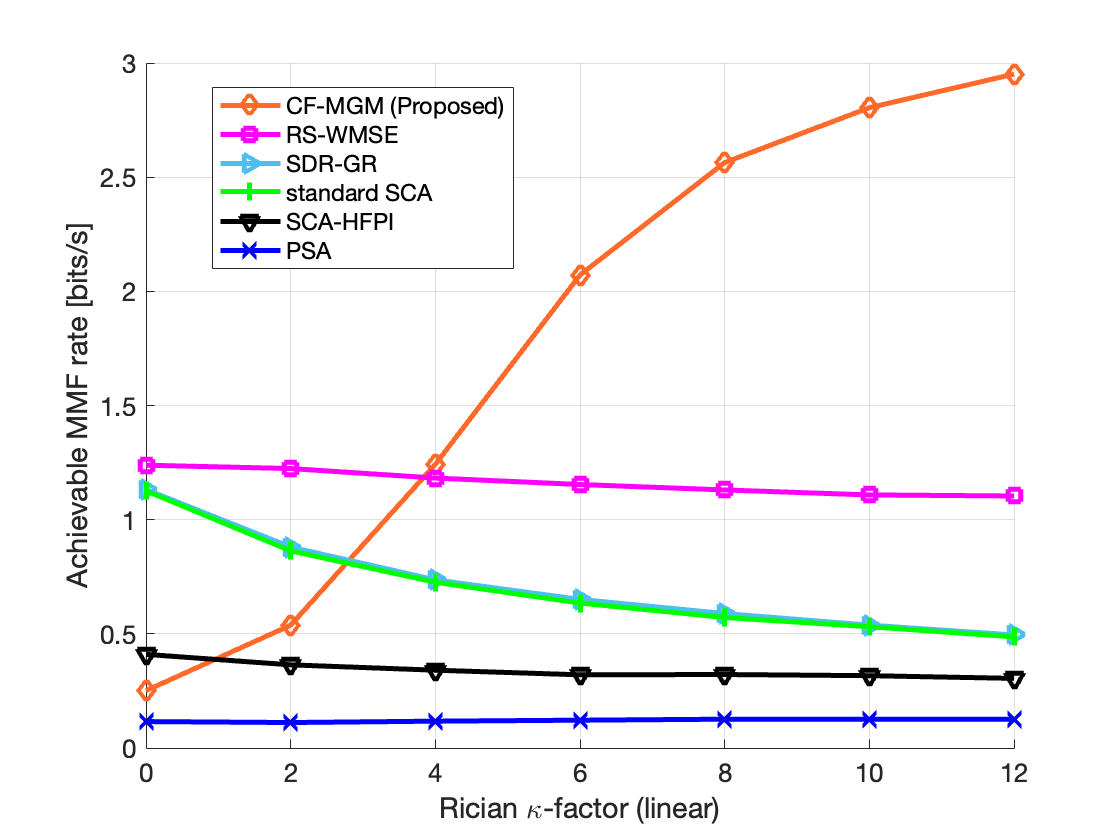}
\caption{Minimum user rate versus Rician-$\kappa$ factor.} \label{kappa}
\end{figure}

\begin{figure}[t]
\centering
\includegraphics[width=0.95\linewidth]{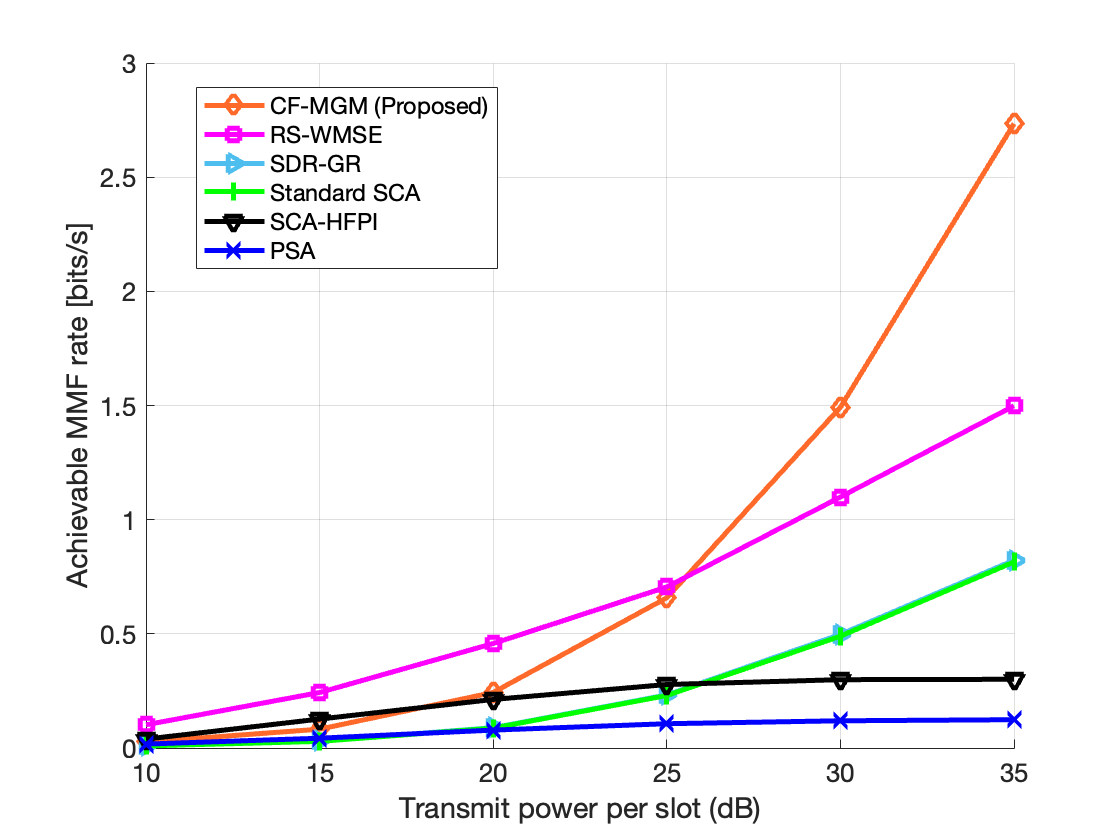}
\caption{Minimum user rate versus Transmit SNR per slot (dB)} \label{PT}
\end{figure}

\begin{figure}[t]
\centering
\includegraphics[width=0.95\linewidth]{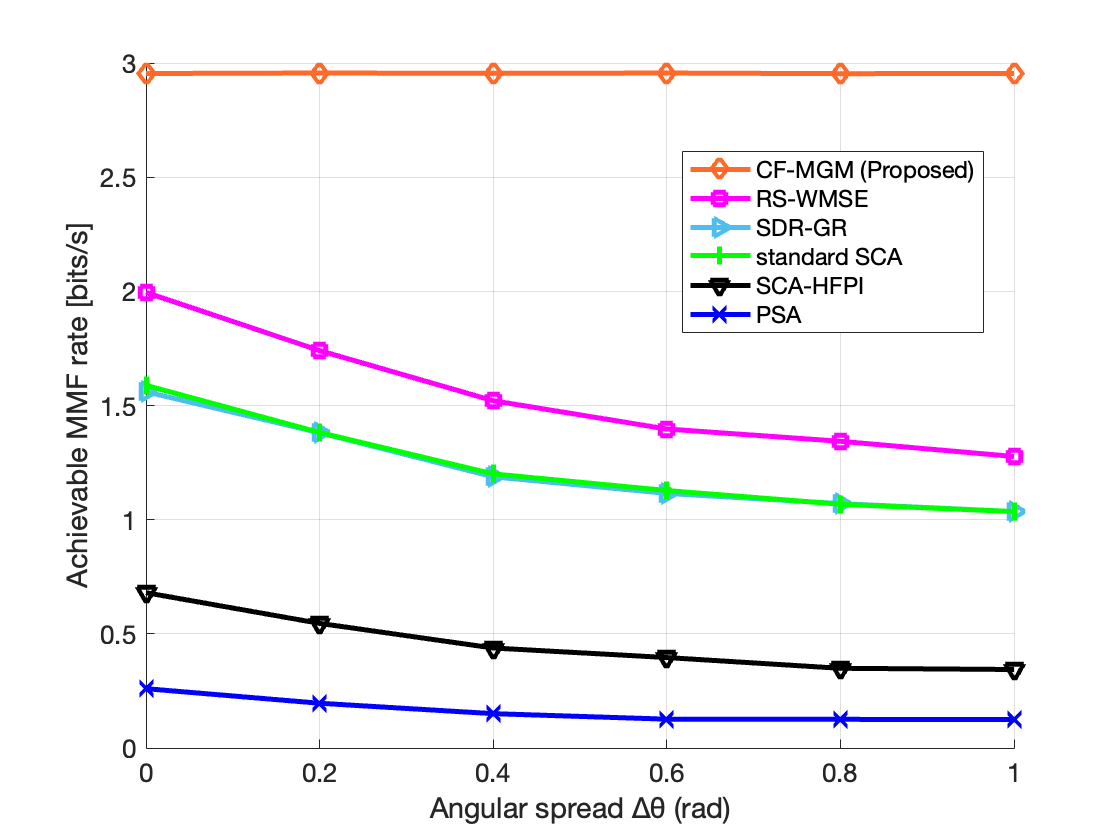}
\caption{Minimum user rate versus AoD angular spread}
\label{fig:AoD}
\end{figure}

Fig.~\ref{UpG} illustrates the minimum user rate as a function of the number of users per group $K_g$. The proposed CF--MGM consistently achieves a pronounced performance over the benchmark schemes, primarily because its transmission and combining structure fully eliminates inter-group interference by construction. Moreover, CF--MGM exhibits significantly stronger robustness as $K_g$ increases. Although a slight performance degradation is observed for very large $K_g$, this loss is mainly attributed to the increased likelihood of cell-edge users experiencing weak large-scale channel gains and the resulting dominance of the additive noise in the MMF bottleneck. Importantly, this degradation is not caused by residual inter-group interference, which is effectively suppressed under the proposed CF--MGM framework.


Fig.~\ref{kappa} shows the minimum user rate as the Rician $\kappa$-factor varies from $0$ to $12$. When $\kappa_{\rm}=0$\, the channel follows Rayleigh fading, whereas at $\kappa=12$, the LoS component accounts for approximately 92\%\ of the total received power, representing a strongly LoS-dominant environment that is typical of practical mmWave systems. For small values of $\kappa$, the intra-group user channels are weakly aligned, which makes it difficult for a single multicast beam to equalize user rates within each group, although the inter-group subspaces remain relatively well separated. As $\kappa$ increases and the LoS component becomes dominant, intra-group alignment improves, but cross-group correlation also increases when angular separation among groups is insufficient. This behavior explains why the benchmark schemes exhibits performance degradation  as $\kappa$ increases: stronger inter-group coupling in LoS-dominant channels directly leads to a more severe interference-limited MMF bottleneck.  In contrast, the proposed CF--MGM exhibits a markedly different trend. In the low-$\kappa$ (scattering-dominant) regime, CF--MGM may yield a slightly lower MMF rate due to imperfect CSIR resulting from quantized receiver-side channel estimation, which limits the accuracy of the linear combining. However, as $\kappa$ increases and the channel becomes more mmWave/LoS-like, CF--MGM increasingly benefits from its CSIT-free multi-slot transmission and combining structure that eliminates inter-group interference by construction. Consequently, its MMF rate improves significantly and becomes substantially higher than the benchmark schemes in the LoS-dominant region, highlighting that the proposed CF--MGM design is particularly well suited to realistic mmWave channels.


Fig.~\ref{PT} depicts the MMF rate as a function of the per-slot transmit power $P_t$, which varies from $10$ to $35$\,dB.
According to~\eqref{eq:SINR_CFMGM}, the post-combining SINR of each user under CF--MGM scales proportionally with the allocated group power $p_g$. Moreover, under the proposed closed-form power allocation, $p_g$ increases linearly with $P_t$. Consequently, CF--MGM achieve a rate growth of $\log_2(1+cP_t)$, for some constant $c$, without exhibiting early saturation. In contrast, all benchmark schemes remain fundamentally interference-limited, even as $P_t$ increases. As a result, their minimum user rates grow only marginally and eventually saturate in the high-SNR regime. This behavior is a direct manifestation of the MMF–DoF collapse predicted by the theoretical analysis: additional transmit power cannot be effectively translated into fairness gains due to persistent inter-group interference. Consequently, the performance gap between CF--MGM and the baseline schemes widens progressively with increasing $P_t$, highlighting the ability of the proposed CSIT-free multi-slot framework to fundamentally overcome the DoF bottleneck of conventional MGM designs.


Fig.~\ref{fig:AoD} illustrates the achievable MMF rate as a function of the intra-group AoD angular spread, where a larger spread indicates that users within the same multicast group are more widely dispersed in the angular domain. As the angular spread increases, all benchmark beamforming-based schemes exhibit a pronounced performance degradation. This is because a single multicast beam becomes increasingly ineffective at simultaneously aligning with all users in the group and suppressing leakage toward other groups, thereby pushing the system into a more severely interference-limited worst-case user SINR regime. In contrast, the proposed CF--MGM framework maintains a nearly constant MMF rate over a wide range of angular spreads. This robustness stems from its structured multi-slot transmission, which eliminate inter-group interference by construction and decouple the MMF performance from the instantaneous angular geometry. Consequently, CF--MGM avoids the rapid saturation behavior observed in conventional schemes under realistic user geometries with large AoD dispersion, further confirming its effectiveness in geometry-limited mmWave multicast scenarios.



\begin{table}[ht]
\caption{{Average CPU Time (ms) vs. Number of Users per Group}}
\setlength{\tabcolsep}{5pt}
\renewcommand{\arraystretch}{1.1}

 \centering

 \begin{tabular}
  {P{40pt} P{40pt} P{40pt} P{40pt} P{40pt} P{40pt}}
   \hline
  $K_g$ & 16 & 24 & 32 & 40 \\ 
\hline
 {\bf CF-MGM} & {$0.079$} & {$0.081$} & {$0.107$} & {$0.140$}  \\
 {\bf PSA} & {$3.970$} & {$4.102$} & {$4.181$} & {$4.442$} \\
 {\bf SCA-HFPI} & {$76.10$} & {$130.8$} & {$182.1$} & {$244.5$} \\ 
\hline
\end{tabular}
\end{table}

Table I reports the average CPU time as a function of the number of users per group. The proposed CF–MGM framework consistently incurs a negligible runtime owing to its closed-form power allocation and the absence of iterative optimization procedures. We omit standard SCA, RS–WMSE, and SDR–GR since their CVX-based implementations incur significantly longer runtimes than the three schemes reported. Among the reported baselines, SCA–HFPI exhibits the largest runtime due to its nested iterative structure with repeated inner fixed-point updates within an outer SCA loop, while PSA is relatively lighter but remains noticeably slower than CF–MGM as it relies on first-order subgradient iterations together with repeated worst-case user identification and projection steps.



\section{Conclusion}

We investigated downlink multi-group multicast (MGM) transmission in overloaded mmWave systems and identified a fundamental limitation of conventional (single-slot) MGM schemes: under realistic LoS-dominant geometries, the MMF-DoF collapses regardless of beamforming optimization. To overcome this geometry-induced bottleneck and the heavy reliance on instantaneous CSIT, we proposed a CSIT-free MGM framework based on a deterministic multi-slot transmission structure. By combining structured precoding with simple receiver-side processing, the proposed CF–MGM eliminates inter-group interference by construction and achieves a strictly positive MMF–DoF in overloaded regimes. By combining structured precoding with simple receiver-side processing, the proposed CF–MGM eliminates inter-group interference by construction and achieves a strictly positive MMF–DoF in overloaded regimes. Overall, CF–MGM provides a scalable and practically attractive paradigm for fairness-aware multicast transmission in future overloaded mmWave networks, where user geometry and CSIT acquisition pose fundamental challenges.

\bibliographystyle{IEEEtran}
\bibliography{CS_REF}



\end{document}